\documentclass[lettersize,journal]{IEEEtran}
\usepackage{graphicx}
\usepackage{dcolumn}
\usepackage{bm}
\usepackage{xcolor}
\usepackage{amsmath,amsfonts,amssymb}
\usepackage{lipsum}

\begin{document}

\title{Microwave Loss in Low-Temperature ICPCVD Silicon Nitride: Dependence on Precursor Ratio and RF Substrate Bias}

\author{%
H. Athreya, J. Cornelison, M. Mikhailov, R.J.H. van der Kolk, C. Chang, T. Cecil, M. N. Ali, J. A. Zebrowski %
\thanks{(Corresponding author: Hrushikesha Athreya)}%
\thanks{H. Athreya is with the Kavli Institute for Cosmological Physics and the University of Chicago, 5640 South Ellis Avenue, Chicago, IL 60637 USA, and also with Argonne National Laboratory, 9700 South Cass Avenue, Lemont, IL 60439 USA (e-mail: athreya7@uchicago.edu).}
\thanks{J. Cornelison was with Argonne National Laboratory, 9700 South Cass Avenue, Lemont, IL 60439 USA, and is now with Fairfield University, 1073 North Benson Road, Fairfield, CT 06824 USA (e-mail: jcornelison@fairfield.edu).}
\thanks{M. Mikhailov, R.J.H. van der Kolk, and M. N. Ali are with the Faculty of Applied Sciences, Delft University of Technology and the Kavli Institute of Nanoscience Delft, Lorentzweg 1, 2628 CJ Delft, The Netherlands (e-mail: m.mykhaylov-1@tudelft.nl, r.j.h.vanderkolk@tudelft.nl, m.n.ali@tudelft.nl).}
\thanks{C. Chang is with Argonne National Laboratory, 9700 South Cass Avenue, Lemont, IL 60439 USA, the University of Chicago and the Kavli Institute for Cosmological Physics, 5640 South Ellis Avenue, Chicago, IL 60637 USA (e-mail: clchang@uchicago.edu).}
\thanks{T. Cecil is with Argonne National Laboratory, 9700 South Cass Avenue, Lemont, IL 60439 USA  (e-mail: cecil@anl.gov).}
\thanks{J. A. Zebrowski is with the Kavli Institute for Cosmological Physics, Enrico Fermi Institute, and the University of Chicago, 5640 South Ellis Avenue, Chicago, IL 60637 USA (e-mail: j.z@uchicago.edu).}}

\maketitle

\begin{abstract}
Low-loss dielectric films are an important component of superconducting circuits with applications ranging from quantum information to superconducting detectors for astrophysics. 
In millimeter-wave detector circuits, dielectric loss can be the limiting factor on efficiency and resonator performance. Some advantageous detector fabrication architectures require dielectric deposition at temperatures substantially lower than those conventionally used to achieve low dielectric loss. 
In this work, we characterize silicon nitride (SiN\(_x\)) films deposited by inductively coupled plasma chemical vapor deposition (ICPCVD) at $150\,^{\circ}\mathrm{C}$ for use in temperature-constrained superconducting detector fabrication. 
We measure the dielectric loss tangent of these films at $\sim 4-7$ GHz using superconducting niobium microstrip resonators and extract the loss by fitting the temperature-dependent resonant frequency to a two-level-system model. 
We vary the \(N_2/SiH_4\) precursor-flow ratio between dielectric depositions and find a clear trend of decreasing loss with decreasing \(N_2/SiH_4\). 
We additionally find that applying an RF substrate bias during deposition reduces the loss compared to films deposited without bias. 
These results identify precursor ratio and substrate bias as practical process parameters for improving low-temperature SiN\(_x\) for superconducting detector circuits.
\end{abstract}


\section{\label{sec:intro}Introduction}

Superconducting circuits play a central role in applications ranging from astrophysical detectors to quantum information processors. 
In these systems, dielectric loss can degrade device performance by attenuating propagating signals and limiting resonator quality factor and optical efficiency.
These effects are particularly important for astrophysical applications, where detector efficiency directly impacts scientific sensitivity. 

One challenge for detector fabrication is that the dielectric deposition process can be constrained by the detector fabrication architecture. 
For example, for on-chip spectrometers for astrophysical applications, radiation propagates down a  superconducting transmission line and is split into frequencies by frequency-selective resonators, which are all impacted by dielectric losses~\cite{2023Slim,2026tifuun}. 
In the conventional fabrication process for these devices, the ground plane is deposited first, then the dielectric, and then the resonator structures are deposited on top. 
Often the dielectric is deposited at elevated temperatures, which generally produces very low-loss films e.g., silicon nitride ($\text{SiN}_{\text{X}}$) deposited in this way can achieve $\tan{\delta} \sim  \mathcal{O}(10^{-5})$\cite{lisovenko2023}. 
However, some detector architectures instead benefit from an inverted fabrication scheme with the resonator structures fabricated before dielectric and ground plane~\cite{2023Slim}. 
For example, this geometry can be advantageous in optically coupled systems to mitigate parasitic coupling.
As a drawback, this inverted process where aluminum is deposited first constrains the temperatures available for subsequent processing because aluminum thin films can form hillocks when exposed to temperatures above $150\,^{\circ}\mathrm{C}$ \cite{ericson1991transmission-b16}.
As a result, standard high-temperature deposition methods that produce very low-loss dielectrics are incompatible with this geometry, while lower-temperature alternatives typically yield higher-loss films.

The dielectric loss of a film depends not only on the deposition temperature but also on the details of the deposition process. In particular, the loss of SiN\(_x\) has been observed to vary with nitrogen content~\cite{lisovenko2023}, motivating a systematic study of the precursor-flow ratio at the lower deposition temperature required for these inverted fabrication architectures. RF substrate bias provides an additional means of modifying SiN\(_x\) film growth. Previous studies have shown that RF bias can change the plasma chemistry and increase ion bombardment of the growing film, resulting in changes in film density, composition, and bonding structure~\cite{dergez2015impact} . 

In this work, we explore SiN\(_x\) films deposited by inductively coupled plasma chemical vapor deposition (ICPCVD) at \(150\,^{\circ}\mathrm{C}\). ICPCVD, unlike classic plasma enhanced chemical vapor deposition (PECVD) can separately tune the plasma density and ion energy~\cite{Lieberman2005}, opening up a much larger parameter space~\cite{Hopwood1992}. It has enabled the deposition of high quality films at much lower temperatures~\cite{Lee2000,Shao2016}. We measure the dielectric loss tangent at approximately \(4\)–\(7\) GHz using superconducting niobium microstrip resonators designed so that the electromagnetic fields are largely confined within the dielectric under test. Between depositions, we vary the \(N_2/SiH_4\) precursor-flow ratio and separately investigate the effect of RF substrate bias. We find a clear trend of decreasing loss with decreasing \(N_2/SiH_4\), together with lower loss in films deposited with RF substrate bias.

\section{Test Device Design and Fabrication}
\label{sec:device_design}

We characterize the dielectric loss using half-wavelength microstrip resonators based on the test structures described in~\cite{Chang2015}. Each test chip consists of four microstrip resonators capacitively coupled to a coplanar waveguide (CPW) readout feedline. In this geometry, nearly all of the electric field energy is confined within the dielectric layer under test, between the superconducting ground plane and the signal line, giving a filling fraction $F \approx 1$. The measured resonator response is therefore highly sensitive to loss in the dielectric. 

\begin{figure}[h] 
    \begin{center}
    \includegraphics[width=0.98\linewidth,trim=0px 0px 0px 0px, clip]{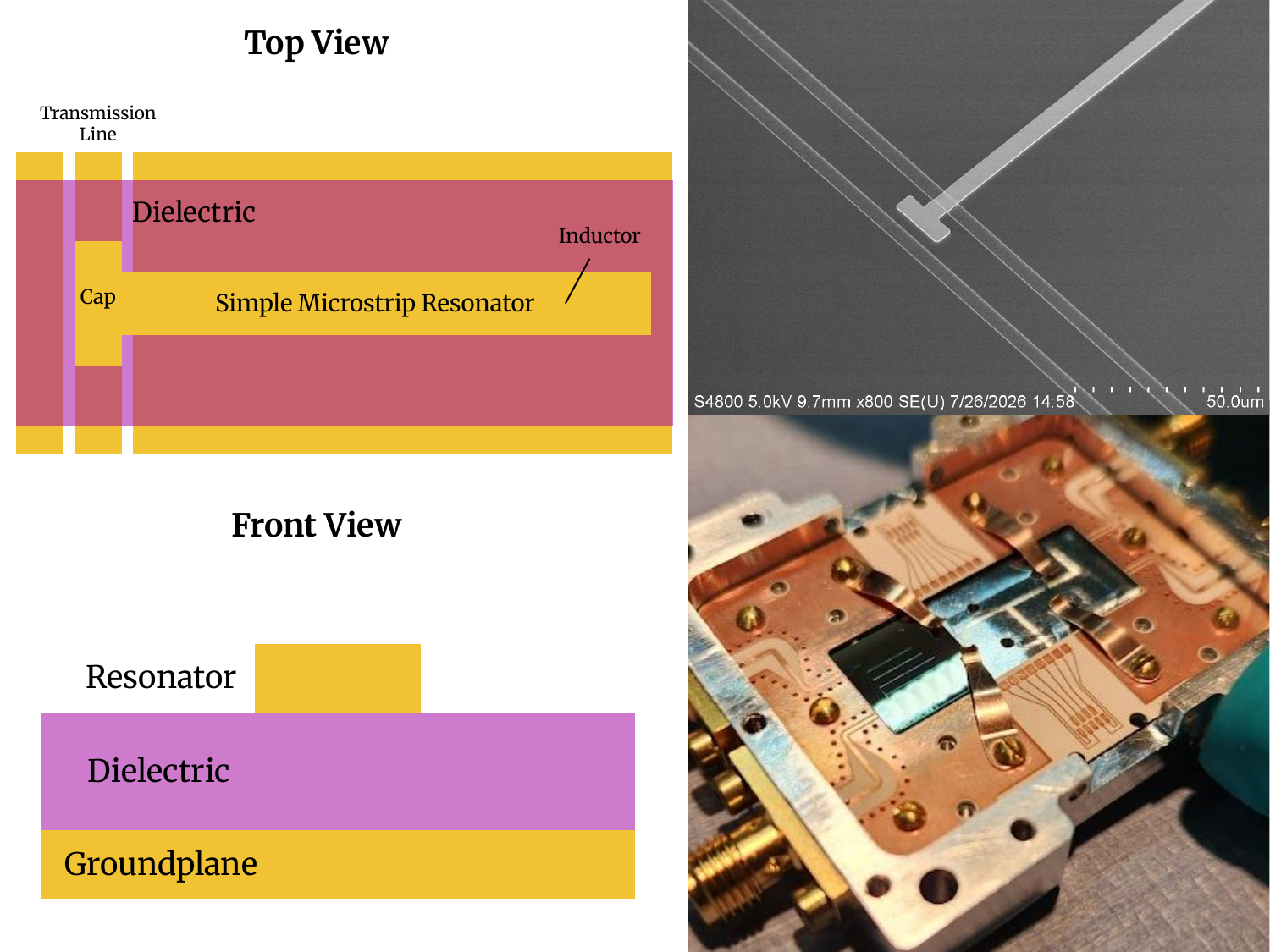}
    \end{center}
    \caption{Top Left: simplified diagram of one microstrip resonator capacitively coupled to a feedline from a top view. Bottom Left: simplified cross section of one microstrip resonator showing fabrication stack-up. Top Right: SEM image of one resonator structure where the resonator microstrip capacitively couples to a transmission line. Bottom Right: test sample in a test box. Each sample was wirebonded to the box's ground and feedline connections.}
    \label{fig:test_device}
\end{figure}

Each four-resonator sample was patterned on a $1\,\mathrm{cm} \times 1\,\mathrm{cm}$ silicon substrate. Resonator lengths range between 7\,mm and 8.5\,mm, yielding resonant frequencies spaced by $\sim 0.3\,\text{GHz}$ and ranging from 4.5\,GHz to 6.5\,GHz depending on the permittivity of the dielectric under test.
A schematic cross section of the microstrip geometry, an SEM micrograph of a representative resonator, and an image of a packaged test chip are shown in Figure~\ref{fig:test_device}.

All samples were fabricated using the same resonator geometry and superconducting materials, with dielectric composition varying between samples. 
Because these resonators are not optically coupled, the inverted-stack process described in Section~\ref{sec:intro} is unnecessary.
The devices were therefore fabricated by first depositing the ground plane, followed by the dielectric and the resonator structures.

We fabricated our devices on high-resistivity \mbox{($>10000~\Omega\cdot\mathrm{cm}$)} silicon wafers. The native oxide was stripped from the substrates using buffered oxide etch (BOE), after which the substrates were rinsed in deionized water and immediately loaded into a UHV chamber for deposition of the ground plane layer. For the ground plane, we deposited 170~nm of niobium using magnetron sputtering and capped it \textit{in situ} with a 30~nm layer of reactively sputtered NbTiN to protect the base layer from oxidation. We patterned the ground plane using e-beam lithography and reactive ion etching in an $\mathrm{SF}_6/\mathrm{O}_2$ plasma. After resist removal and cleaning, we proceeded with dielectric deposition. The dielectric films were deposited by ICPCVD at $150^\circ\mathrm{C}$, consistent with the temperature constraint described in Section~\ref{sec:intro}. We used an Oxford Instruments PlasmaPro100 ICPCVD deposition system suitable for processing wafers up to 8~inches in diameter. SiN$_x$ films were deposited using SiH$_4$ and N$_2$ as precursor gases at a pressure of 2~mTorr and an ICP power of 500~W. Hydrogenated amorphous silicon (a-Si:H) films were deposited using SiH$_4$ as the precursor gas and Ar as the carrier gas at a pressure of 5~mTorr and an ICP power of 1000~W. The thickness of all dielectric films was 300~nm.

Across the primary sample series, we varied the N$_2$ and SiH$_4$ precursor flow rates to change the \(N_2/SiH_4\) precursor-flow ratio while holding the remaining deposition parameters fixed. A second set of samples was deposited under comparable precursor-flow conditions with a 5~W RF substrate bias. The refractive index of each film was measured at 632.8 nm and is reported in Table \ref{tab:table1}.  While refractive index is not a direct measurement of stoichiometry, under comparable deposition conditions, higher refractive index provides a qualitative indication of more silicon-rich films~\cite{verlaan2009effect, sun2024cryogenic}.

The residual stress of the dielectric films was measured using the wafer curvature method with a commercial laser-scanning system. The biased films exhibited much larger compressive stress as seen in Table \ref{tab:table1}, with measured values of approximately \(-640\) to \(-730\) MPa compared with approximately \(-280\) to \(-300\) MPa for the corresponding unbiased films. This increase in stress with applied RF power is consistent with the work of Dergez et al. \cite{dergez2015impact}, who found that even low RF power led to a changed plasma chemistry, inhibited the formation of N--H bonds during film growth, and produced a denser film with higher compressive stress. We also note that the deposition recipe of Paik et al.~\cite{Paik2010}, which produced low-loss SiN\(_x\) at \(300\,^{\circ}\mathrm{C}\), also included a low (4~W) nominal RF power, although that work did not compare films grown with and without it. The deposition conditions for each sample are summarized in Table~\ref{tab:table1}. After dielectric deposition, the microstrip resonators were patterned by liftoff using a bilayer e-beam resist mask, followed by e-beam evaporation of 400~nm of niobium. No intentional substrate heating was used during either magnetron sputtering or e-beam evaporation.

\begin{table*}[!ht]
\caption{Dielectric deposition properties and measured loss tangents\label{tab:table1}}
\renewcommand{\arraystretch}{1.2}
\centering
\begin{tabular}{|c|c|c|c|c|c|c|}
\hline
Material & $\text{N}_{2}$ Flow Rate (sccm) & $\text{SiH}_{4}$ Flow Rate (sccm) & Stress (MPa)& n (632.8 nm) & $\tan\delta$ & $\sigma_{\tan\delta}$ \\
\hline
$\text{SiN}_{\text{X}}$ & 30 & 21 & -48 & 1.79 & $3.30\times10^{-3}$ & $3.3\times10^{-5}$ \\
\hline
$\text{SiN}_{\text{X}}$ & 25 & 26 & -283 & 2.10 & $9.14\times10^{-4}$ & $9.5\times10^{-6}$ \\
\hline
$\text{SiN}_{\text{X}}$ + RF bias & 25 & 26 & -734 & 2.06 & $3.68\times10^{-4}$ & $4.7\times10^{-6}$ \\
\hline
$\text{SiN}_{\text{X}}$ & 21 & 30 & -294 & 2.50 & $4.68\times10^{-4}$ & $6.6\times10^{-6}$ \\
\hline
$\text{SiN}_{\text{X}}$ + RF bias & 21 & 30 & -638 & 2.40 & $1.04\times10^{-4}$ & $8.2\times10^{-6}$ \\
\hline
a-Si:H & 0 & 20 & -60 & 3.30 & $6.04\times10^{-5}$ & $3.7\times10^{-6}$ \\
\hline

\end{tabular}\\[5pt]
\begin{minipage}{0.85\textwidth}
\footnotesize
\textit{Note:} Deposition conditions and measured properties of the dielectric films, including precursor flow rates, residual stress, refractive index at 632.8\,nm, and microwave loss tangent. $\sigma_{\tan\delta}$ denotes the $1\sigma$ standard deviation across resonators for each deposition recipe.
\end{minipage}
\end{table*}

\section{Experimental Procedure}
\label{sec:exp_procedure}
Each sample was wirebonded inside a light-tight test box, shown in Figure~\ref{fig:test_device}, and mounted onto the cold stage of an adiabatic demagnetization refrigerator (ADR) sitting at a base temperature of 0.1\,K. An $S_{21}$ forward transmission of each sample was measured with a vector network analyzer (VNA) over 4--7\,GHz, and resonances were identified as sharp dips in the $|S_{21}|$ frequency sweep. The set of potential resonances were confirmed by their characteristic response to sweeps in power and temperature. The resonant frequency of each resonator was tracked as a function of temperature, and the resulting fractional frequency shift was fit to a TLS model to extract the dielectric loss tangent, as described below.

\subsection{Dielectric loss from TLS}
The two-level system (TLS) model describes the behavior of atoms tunneling between two distinct energy states at extremely low temperatures\cite{Phillips1987}.
Losses in a dielectric due to TLS are well modeled assuming a log-uniform distribution of tunneling states, resulting in a temperature-dependent shift in resonant frequency for a given resonator\cite{Gao2008}:

\begin{equation}\label{eq:tsweepeq}
    \frac{f\left(T\right)-f_0}{f_0} = \frac{F \tan\delta}{\pi}\left[\text{Re}\Psi\left(\frac{1}{2}-\frac{\hbar f_0}{j k_B T}\right)-\log\frac{\hbar f_0}{k_B T}\right]
\end{equation}

where $f$ is the resonant frequency at temperature $T$, $f_0$ is the $T=0$ resonant frequency, $F$ is the filling factor, $\tan\delta$ is the dielectric loss, $\Psi$ is the digamma function, $\hbar$ is the reduced Planck constant, and $k_B$ is the Boltzmann constant.

We constrained $\tan\delta$ by tracking the resonant frequency of our resonator through a range of temperatures and fitting to the model in Eq. \ref{eq:tsweepeq}.
Measurement temperatures were bounded at $\sim 0.1\,\text{K}$ by the base temperature of our ADR and $\sim 1.5\,\text{K}$ where frequency shifting becomes dominated by kinetic inductance changes induced by Cooper-pair-breaking.
The filling fraction is determined primarily by the geometry of the devices which was designed to be $\approx1$.

\begin{figure}[h] 
    \centering\includegraphics[width=0.99\linewidth,trim=10px 10px 10px 10px, clip]{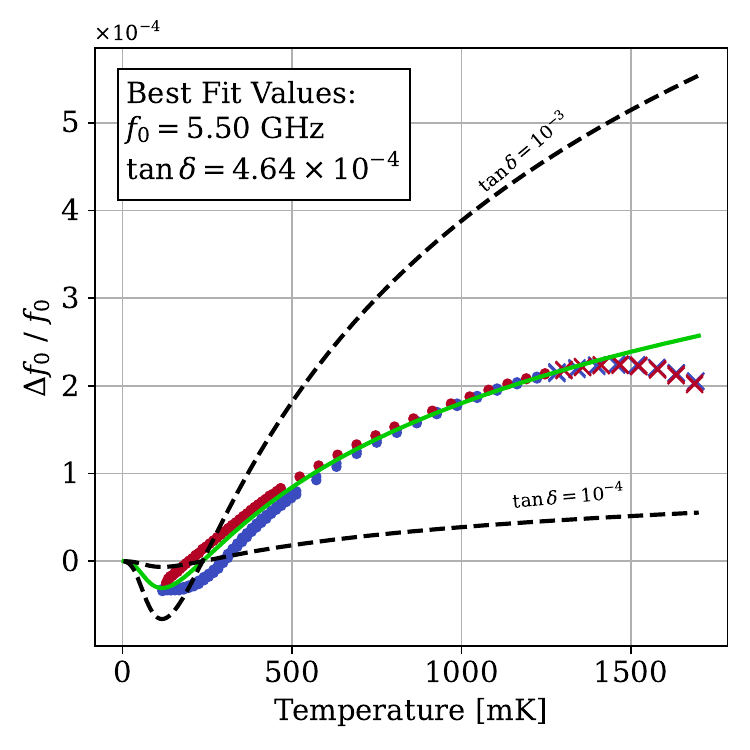}
    \caption{Example of fractional frequency shift vs. temperature data of a SiN resonator sample. Data in blue are measured as the sample temperature was ramped up and red data were measured when the sample temperature was ramped downward. The speed of the ramp results in a measurable hysteresis, but has negligible impact on loss measurements. The \texttt{o} symbols are measured data included in the fits while \texttt{x} denotes measured data that was excluded from the fits due to the domination of Cooper-pair-breaking effects at higher temperatures. The green solid line is the best-fit model.}
    \label{fig:TLSfig}
\end{figure}

An example measurement is shown in Figure~\ref{fig:TLSfig}.
Temperature sweeps were performed by continuously ramping the ADR's magnet current both upward to $1.5\,\text{K}$ and then downward to a base temp of $\sim 0.1\,\text{K}$, taking $15$ minutes in each direction.
The speed at which the magnet was ramped contributes to some hysteresis in the frequency-shift measurements, which can be seen in Figure~\ref{fig:TLSfig}, with slower sweeps possessing less hysteresis.
We found, however, the impact on $\tan\delta$ measurements between our standard 15-minute sweeps and those of $\sim 1$ hour to be $< 3$\%, so long as upward and downward temperature sweeps were included in the fits.

\subsection{Loss Tangent Uncertainty}
Each resonator on a sample provides an independent measurement of $\tan\delta$. The loss tangent results of a given film are reported as the mean of the individual resonator fits, with the uncertainty given as the $1\sigma$ sample standard deviation across those resonators. For deposition recipes measured on more than one sample, resonators from all samples of that recipe were combined before computing the mean and standard deviation. Between three and four resonators contributed to each value, depending on the number of resonators successfully identified for each recipe. The resulting loss tangents and uncertainties for all samples are summarized in Table~\ref{tab:table1}.

\section{Results \& Discussion}
\label{sec:results}

\begin{figure*}[!t]
   \centering\includegraphics[width=\textwidth,trim=4px 10px 8px 7px, clip]{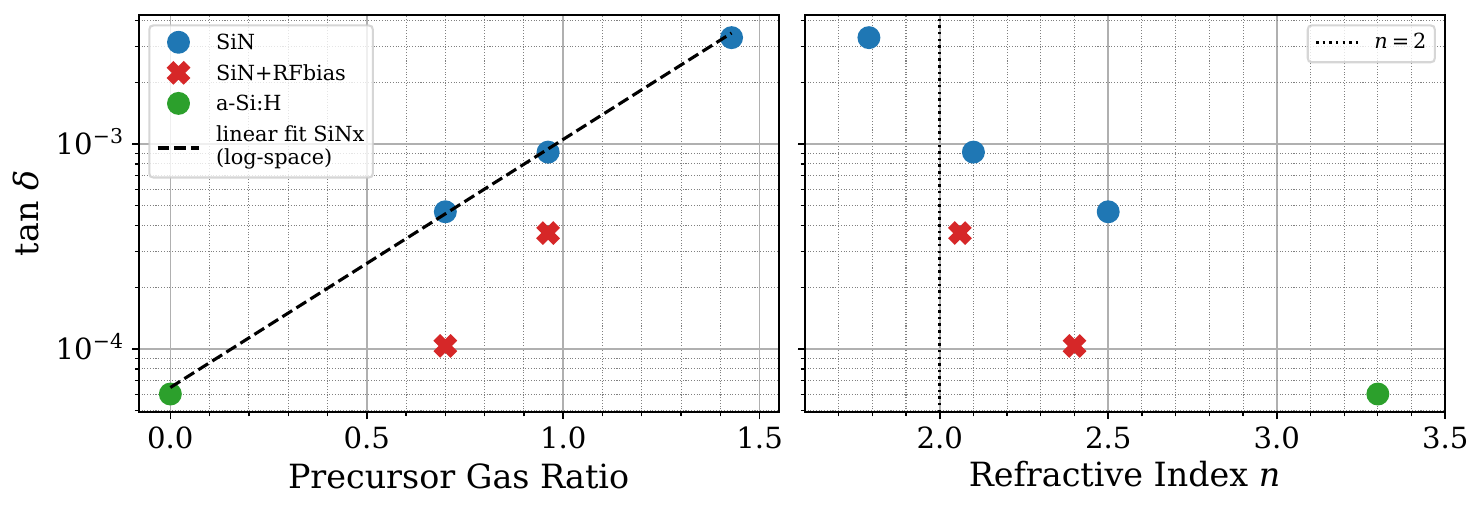}
    \caption{Measured $\tan\delta$ for each deposition recipe in Table~\ref{tab:table1}, all deposited by ICPCVD at $150\,^\circ$C, plotted against $\mathrm{N_2/SiH_4}$ precursor gas ratio (left) and refractive index $n$ measured at 632.8\,nm (right). Left: the dashed line shows a linear fit in log space to the unbiased SiN$_x$ samples. Extrapolating this trend down to a flow ratio of zero shows agreement with our independently measured a-Si:H sample, which was not included in the fit. Right: loss decreases as $n$ increases, consistent with the films becoming more silicon-rich. The dotted line marks $n=2$, which marks an approximate reference value for stoichiometric SiN\(_x\). In both panels, SiN$_x$ samples deposited with a 5\,W RF substrate bias show lower loss compared to unbiased samples at the same gas ratio, despite only modest changes in $n$. Uncertainties for each recipe are listed in Table~\ref{tab:table1}.}
    \label{fig:loss_vs_flow}
\end{figure*}

\begin{figure}[ht] 
    \centering\includegraphics[width=\linewidth,trim=0px 0px 0px 0px, clip]{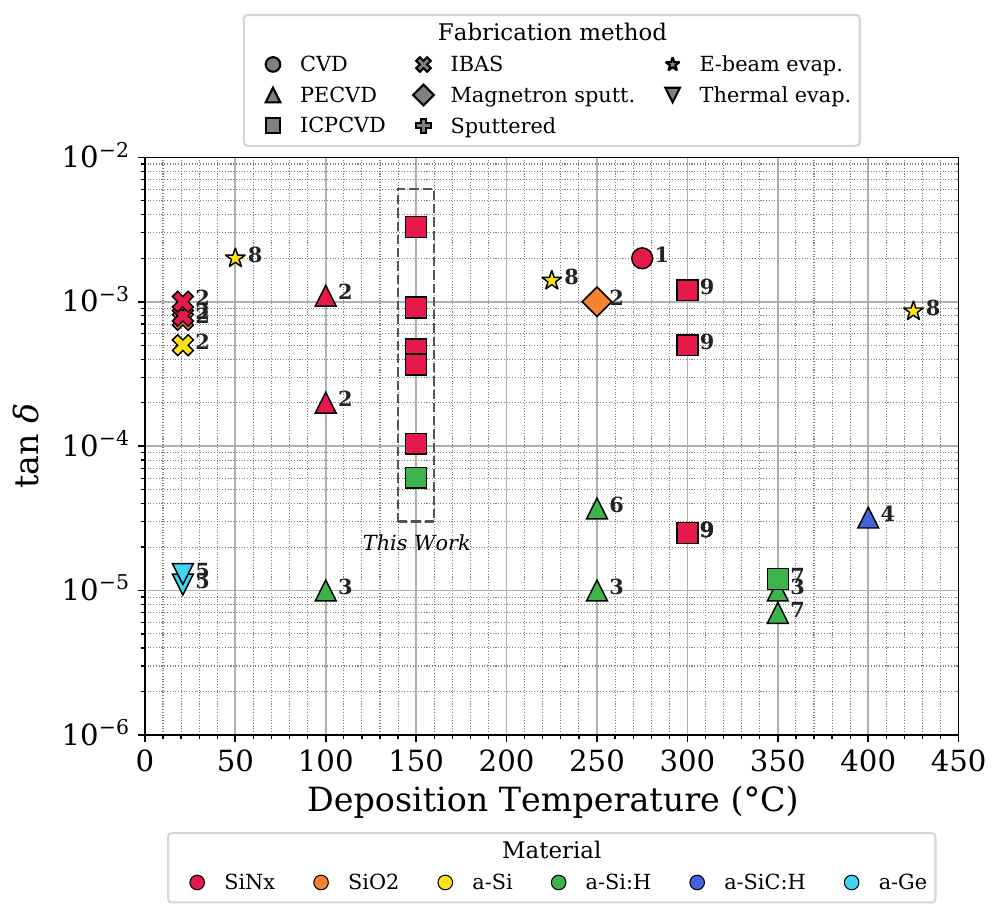}
    \caption{Comparison of $\tan\delta$ measured in this work (dashed box at $150\,^\circ$C) with published microwave-frequency loss tangents of deposited dielectrics with respect to the deposition temperature. Marker shape indicates deposition method while color indicates dielectric material. 1 is from \cite{Pan2023}, 2 is from \cite{lisovenko2023}, 3 is from \cite{Buijtendorp2022a}, 4 is from \cite{Buijtendorp2022b}, 5 is from \cite{Kopas2021}, 6 is from \cite{Hahnle2021}, 7 is from \cite{Defrance2024}, 8 is from \cite{MolinaRuiz2021}, 9 is from \cite{Paik2010}}
    \label{fig:loss_vs_Tsub}
\end{figure}

Figure~\ref{fig:loss_vs_flow} plots the dielectric loss tangents extracted across our set of samples as a function of the \(N_2/SiH_4\) precursor gas ratio and refractive index summarized in Table~\ref{tab:table1}. The dominant loss mechanism at the frequencies and temperatures used in these measurements is resonant absorption by two-level systems (TLS), as described in Section~\ref{sec:exp_procedure}.
 
We observe a linear relationship in log-space for our $\text{SiN}_{\text{X}}$ samples between the $\text{N}_2$-to-$\text{SiH}_4$ precursor gas flow ratio and their measured loss tangent. The measured refractive index increases systematically from \(n=1.79\) to \(n=2.50\) as the \(N_2/SiH_4\) ratio increases, consistent with the films becoming progressively more silicon rich. This spans a broad range of refractive indices, indicative of a substantial range of film compositions. This loss trend extends down to our a-Si:H sample, suggesting a continuous dependence of TLS loss on film composition rather than a sharp stoichiometric transition. This behavior differs from previous work with dielectrics deposited at $300\,^\circ\text{C}$~\cite{Paik2010}, which reported a sharp drop in $\tan\delta$ from $\sim\!10^{-3}$ to $\sim\!10^{-5}$ when varying the precursor ratio from $\sim\!1.2$ to $\sim\!1.1$, attributed to a transition from nitrogen-rich to silicon-rich films. Since refractive index is an indirect measurement of composition, direct compositional measurements would be needed to quantitatively compare the precise Si:N ratio of each film. The absence of the sharp transition in our data and the differences between the trends observed here and in~\cite{Paik2010} may reflect differences in the deposition temperature, the fixed $\text{SiH}_4$ flow rate used in~\cite{Paik2010}, or both.
 
We also find that applying an RF substrate bias during the ICPCVD deposition step reduces the measured loss tangent in our $\text{SiN}_{\text{X}}$ samples. At fixed precursor-flow ratio, the RF bias produces only modest changes to the refractive index, despite a substantial change in the measured dielectric loss. This implies that the reduction in loss is not primarily driven by a large change in overall film composition.  The RF-biased films exhibit substantially larger compressive stress, consistent with previous studies \cite{dergez2015impact} showing that low auxiliary RF power during ICPCVD inhibits N--H bond formation and produces denser, more compressively stressed films. Because N--H related defects have been identified as a likely source of TLS loss in $\text{SiN}_{\text{X}}$ \cite{Paik2010}, an inhibition of N--H bonds from an applied RF bias offers a plausible explanation for the loss we observe. These comparisons suggest that RF bias is a promising addition to a deposition process for reducing loss, although for our samples, additional measurements, e.g. FT-IR, will be required to directly characterize the bonding structure of our films.
 
Figure~\ref{fig:loss_vs_Tsub} places our results in the context of previously published dielectric loss measurements as a function of deposition temperature.
Our $150\,^\circ\text{C}$ ICPCVD-deposited films, in particular the samples deposited with increased RF-bias, represent some of the lowest $\tan\delta$ values reported for dielectrics deposited at this temperature, demonstrating achievable science-grade films compatible with an inverted fabrication scheme.

\section{Conclusion}
In this paper, we characterize the microwave loss of SiN$_x$ films deposited by ICPCVD at $150\,^\circ$C, a temperature that is compatible with inverted-stack detector architectures in which aluminum is deposited before the dielectric. ICPCVD is well suited to this temperature constraint because it decouples plasma density from ion energy, allowing film growth to be tuned without raising the deposition temperature. The loss tangent of our unbiased samples decreases log-linearly with decreasing \(N_2/SiH_4\) precursor gas ratio spanning a broad range of refractive indices, differing from the sharp drop reported by \cite{Paik2010} for dielectrics deposited at $300\,^\circ$C. Applying an RF substrate bias during deposition further reduces the measured loss tangent, allowing us to reach $\tan\delta = 1.04\times10^{-4}$. The accompanying increase in compressive film stress is consistent with RF bias inhibiting N--H bond formation~\cite{dergez2015impact}, a likely source of TLS loss in SiN$_x$. These values are among the lowest reported for SiN$_x$ deposited at or below $150\,^\circ\mathrm{C}$, demonstrating that low-loss films are achievable within the thermal budget of inverted-stack fabrication.

\section*{Acknowledgments}
This work was supported by The Kavli Foundation through a Kavli Institute Collaboration Kickstarter (KICK) grant to the Kavli Institute for Cosmological Physics and the Kavli Institute of Nanoscience Delft.
Work performed at the Center for Nanoscale Materials, a U.S. Department of Energy Office of Science User Facility, was supported by the U.S. DOE, Office of Basic Energy Sciences, under Contract No. DE-AC02-06CH11357.
The authors used generative AI tools, including Anthropic Claude 4.7 Opus and OpenAI ChatGPT 5.5, as interactive assistants in support of literature organization, manuscript and LaTeX preparation, coding assistance, and figure-workflow development. 
The authors guided the use of these tools and reviewed, verified, and revised any AI-assisted material incorporated into the manuscript. 
All scientific judgments, derivations, calculations, interpretations, conclusions, and responsibility for the final content of the manuscript remain solely with the authors.

\bibliographystyle{IEEEtran}
\bibliography{references.bib}

\end{document}